# Combinatorial nanoparticle patterns assembled by photovoltaic optoelectronic tweezers

Carlos Sebastián-Vicente,[1] Pablo Remacha-Sanz,[1] Eva Elizechea-López,[1] Ángel García-Cabañes,[1,2] and Mercedes Carrascosa[1,2,a]

[1]*Departamento de Física de Materiales, Facultad de Ciencias, Universidad Autónoma de Madrid, c/ Francisco Tomás y Valiente 7, 28049 Madrid, Spain*

[2]*Instituto de Ciencia de Materiales Nicolás Cabrera, Universidad Autónoma de Madrid, 28049 Madrid, Spain*

[a]*Author to whom correspondence should be addressed. E-mail:* m.carrascosa@uam.es

Photovoltaic optoelectronic tweezers (PVOT) have been proven to be an efficient tool for the manipulation and massive assembly of micro/nano-objects. The technique relies on the strong electric fields produced by certain ferroelectric materials upon illumination due to the bulk photovoltaic effect (customarily $LiNbO_3$:Fe). Despite the rapid development of PVOT and the achievement of high-quality 1D and 2D particle patterning, research efforts aimed at the fabrication of combinatorial structures made up of multiple types of particles have been scarce. Here, we have established the working principles of three different methods to tackle this pending challenge. To that end, dielectrophoresis and/or electrophoresis acting on neutral and charged particles respectively, have been suitably exploited. Simple mixed structures combining metallic and dielectric nanoparticles of different sizes have been successfully obtained. The results lay the groundwork for future fabrication of more complex combinatorial structures by PVOT, where micro/nanoparticles are the basic building blocks of miniaturized functional devices.

Optical and optoelectronic techniques devoted to the manipulation and assembly of micro- and nano-objects are currently acquiring increasing relevance in nano- and biotechnology. The accurate arrangement of nanoparticles (NPs) in desired patterns has attracted major attention over the past few decades, with myriads of potential applications for sensing, optical and electronic devices, electrochemistry or energy conversion, exploiting the unique properties of NPs.[1-3] Currently, intense research activity is being devoted to this active field, with a particular focus on combining different types of particles in a single pattern. Such structures are usually known as "combinatorial" patterns.[4] So far, the fabrication of combinatorial structures has been explored by several different techniques, each of them with its own advantages, drawbacks and limitations.[4-11] Combinatorial patterning enriches the complexity and functionalities of NP assemblies, opening new routes for multiple applications, such as combinatorial chemistry on a chip, multimodal sensing, NP separation, analysis of particle mixtures or smart encryption for security, to name a few.[12-15]

Photovoltaic optoelectronic tweezers (PVOT) are a noteworthy tool that has undergone strong progress over the last years with remarkable results and widespread potential applications.[16,17] It is based on the electric fields generated upon light excitation of certain noncentrosymmetric materials (usually doped ferroelectrics) via the bulk photovoltaic (PV) effect, mainly iron-doped $LiNbO_3$.[18,19] In $LiNbO_3$:Fe crystals, the light induces directional excitation of electrons from the iron impurities ($Fe^{2+}$) along the polar axis. Then, the electrons migrate until they are trapped in suitable acceptors ($Fe^{3+}$). This process gives rise to charge separation, hence producing an electric field that extends outside the crystal in the form of evanescent fields near its surface (the PV field responsible for particle manipulation and trapping). The PVOT technique stands out for its simplicity and versatility, allowing the electrostatic manipulation of a wide range of micro/nano-objects solely driven by light, without electrodes or external power supplies.

Probably, the main consolidated achievement of PVOT is the flexible fabrication of 1D and 2D micro- and NP trapping patterns. For this purpose, the method has key advantages over other techniques: i) At difference with optical tweezers, it traps in parallel myriads of particles according to the illumination patterns and using low light intensities.[20] ii) Compared with other optoelectronic methods, no electrodes or power supplies are necessary and the electric fields remain after illumination thanks to the low dark conductivity of the crystals, thus giving stability to the patterns.[21-23] Virtually, any kind of particles, charged or neutral, as well as dielectric, metallic or even bio-particles can be organized.[24-27] Similarly, patterning of liquid droplets has also been successfully performed.[28] Moreover, elongated bio-objects (bacteria) and particles have been oriented on the substrate parallel to the electric field.[29,30] All these achievements and properties suggest a large span of possibilities for the application of 1D and 2D micro and nanostructures fabricated by massive bottom-up assembly of NPs. In fact, some of them, such as the fabrication of diffractive optical components or plasmonic structures, have been recently reported.[31,32] However, except for a single result in an early paper,[24] all the patterns produced so far by PVOT are fabricated with only one kind of particles. Meanwhile, the possibility of assembling multiple types of particles in a single combinatorial pattern has already been explored by means of other patterning techniques, as already mentioned.[4] Therefore, a step forward for PVOT will be to combine different kinds of particles, thereby enhancing the functionality of the patterns and enriching their applications.

In this work, we face this challenge and develop several strategies to fabricate NP patterns with different kinds of particles, spatially separated. Metallic and dielectric particles have been used and combined in the same mixed pattern. Different methods based on the multiple possibilities and versatility of PVOT are applied.

So far, in order to fabricate patterns made up of a single type of particle, a two-step method (that we called sequential method) is usually applied.[26,33] First, the PV substrate (LiNbO$_3$:Fe) is illuminated with a light pattern or a moving focused beam, generating an electric field distribution that remains after illumination. Second, the particles are approached to the substrate usually suspended in a nonpolar liquid, where they become trapped according to the electric field pattern. If the particles are charged, they undergo electrophoretic (Coulombian) forces whereas if they are neutral they experience dielectrophoretic forces.[16] Two orientations of the LiNbO$_3$:Fe substrates are used, namely *x*-cut and *z*-cut substrates with the polar axis parallel and perpendicular to the active surface, respectively (see Fig. 1(a)).

Moreover, it has been recently reported that particles previously trapped can be removed from the pattern by illuminating it with a light beam.[34,35] This de-trapping process is due to a charge transfer between the charged LiNbO$_3$:Fe surface and the NPs, which get charged with the same sign. Consequently, the NPs are repelled from the surface. This phenomenon can be used for pattern erasure and reconfiguration.[34] In fact, the mix of trapping and detrapping has been used in the second combinatorial approach described below. Moreover, we have also exploited the charge-transfer phenomenon to obtain charged particles with both signs, which is used in the last method to assemble binary structures comprising oppositely charged particles. To the best of our knowledge, this method for obtaining charged particles had not been applied thus far.

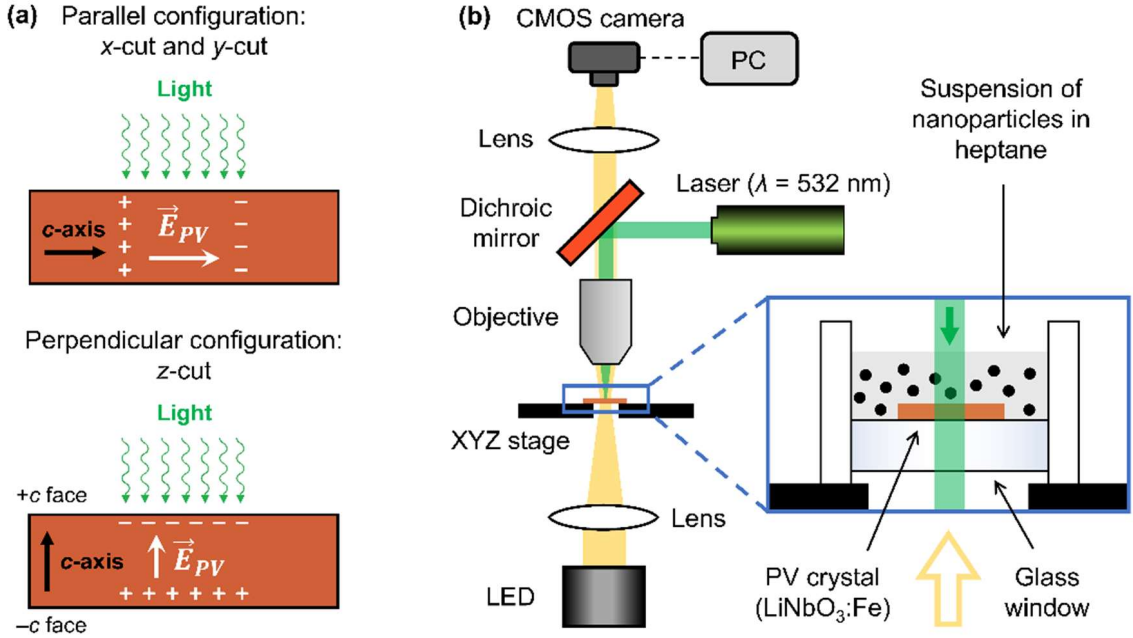

FIG. 1. (a) Ferroelectric crystal orientations available for PVOT. The polar *c*-axis is parallel to the substrate surface in *x*-cut and *y*-cut crystals, as opposed to *z*-cut crystals where it is perpendicular. (b) Schematic diagram of the experimental setup.

As already mentioned, we have used highly-doped $LiNbO_3$:Fe monodomain crystals as PV substrates (0.25 mol% iron concentration), with the active surface perpendicular (*z*-cut) or parallel (*x*-cut) to the polar ferroelectric axis. The experimental setup, depicted in Fig. 1(b), allows light excitation of the PV crystal by a focused Gaussian beam (static or following arbitrary trajectories), as well as real-time visualization of the substrate surface. The wavelength of the excitation laser source is 532 nm, the $1/e^2$ diameter of the Gaussian spot is $2w$ = 150 μm and the typical optical intensity is $I$ = 6 W/cm$^2$ (unless otherwise specified). Dielectrophoretic/electrophoretic manipulation[26] of particles is achieved by placing the substrate inside a cuvette filled successively with different suspensions, as illustrated in the inset of Fig. 1(b). The suspensions are prepared by dispersing particles in a nonpolar medium (heptane) to avoid screening of the evanescent PV fields. Concentrations ranging between 50 mg/L and 300 mg/L have been employed. The particles employed throughout this work are metallic NPs (Ag with average diameter $d$ = 100 nm and Al with $d$ = 70 nm) and dielectric ones ($SiO_2$, $BaTiO_3$ and $Al_2O_3$ with $d$ = 500 nm, 400 nm and 1 μm, respectively).

In this paper, three alternative approaches have been attempted to produce combinatorial NP structures, presented hereinafter.

In the first approach, we use the typical sequential method to fabricate the patterns with PVOT, but iteratively repeating it on the same $z$-cut substrate for different kinds of neutral particles. In the presence of an electric field, neutral particles are polarized, i.e. an electric dipole is induced. As a result, the particles experience a net electric force, widely known as "dielectrophoretic" force. Such forces can be written as a function of the PV electric field as follows:[36]

$$\mathbf{F_{DEP}} = (\mathbf{p} \cdot \nabla) \cdot \mathbf{E_{PV}} \qquad (1)$$

where $\mathbf{p}$ is the induced dipole and $\mathbf{E_{PV}}$ is the PV electric field. In general, the components of the induced dipole may be expressed as $p_i = \alpha_{ij}E_j$, where $\alpha_{ij}$ are the components of the polarizability tensor of the particle embedded in a given surrounding medium. For isotropic spherical particles, the polarizability becomes a scalar, and the dielectrophoretic force can be written as:

$$\mathbf{F_{DEP}} = \frac{1}{2}\alpha \nabla \mathbf{E_{PV}^2} \qquad (2)$$

According to Equation (2), the particles can be directed towards the regions of highest or lowest electric field intensity, depending on whether $\alpha>0$ (positive dielectrophoresis) or $\alpha<0$ (negative dielectrophoresis). In this work, all the particles are more polarizable than the surrounding medium (heptane), and so, positive dielectrophoresis takes place, i.e. they are attracted to the regions of highest electric field.

By using this iterative method, structures with two and three types of particles have been generated. In Fig. 2 we illustrate the result for an H-shaped mixed pattern combining neutral BaTiO$_3$, SiO$_2$ and Ag particles, each of them forming the horizontal, left vertical and right vertical segments, respectively. The particles were trapped on the substrate in that order. The insets show a magnification of small regions from the three segments where one can distinguish the different aspect of each nano-material.

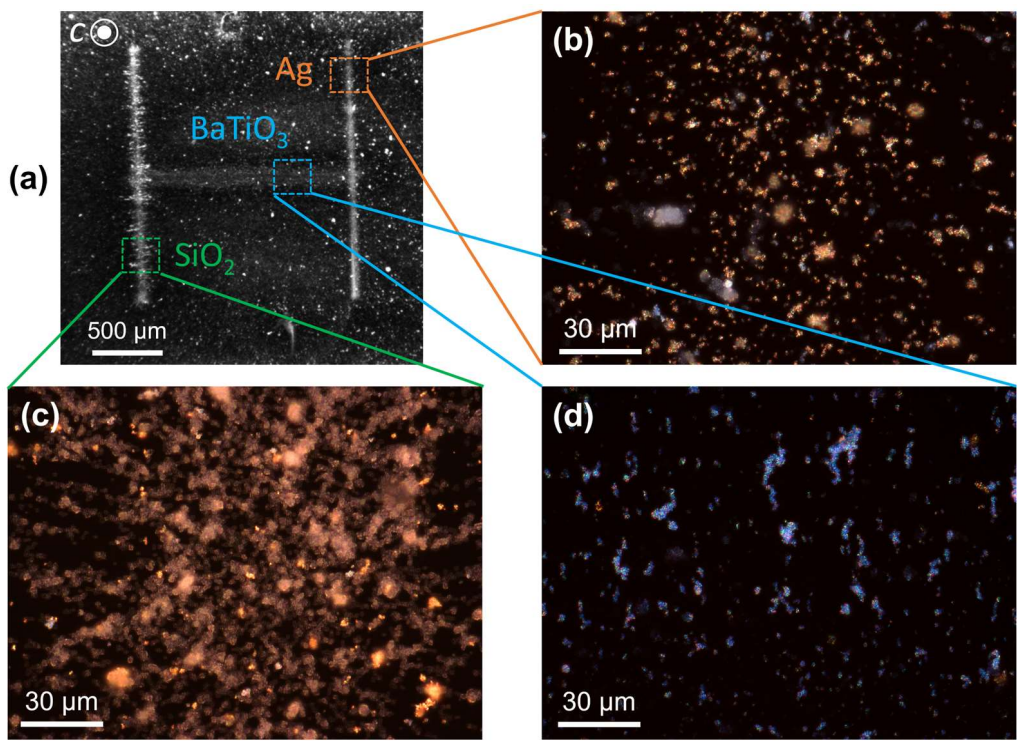

FIG. 2. (a) Image of a combinatorial H-shaped pattern made up of Ag, SiO$_2$ and BaTiO$_3$ NPs (see text for details) (b)-(d) Magnified microphotographs of a small region of each segment of the pattern as indicated in the figure (taken with crossed polarizers).

Overall, the microscope images confirm that different particles predominate in each segment, although a slight cross-contamination may be discerned by careful inspection. This weak but detrimental effect is mainly attributed to unscreened residual electric fields in the background and particle-substrate adhesion forces, leading to some undesired trapping at non-illuminated areas. Although this method has been tested with up to three types of particles so far, in principle the procedure could be iterated an arbitrary number of times. Nevertheless, the aforementioned cross-contamination hinders the scalability of the method in that direction, unless a trapping protocol that mitigates this issue is implemented.

The second proposed strategy is based on light trapping and detrapping and the fabrication process involves three steps, further illustrated in Fig. 3(a):

1) Sedimentation of a homogeneous layer of the first kind of particles (BaTiO$_3$) on a $z$-cut substrate by introducing it in a suspension of those particles (for around 10 min).

2) Erasure of a part of this uniform layer by illumination with a focused light beam and taking advantage of the particle ejection effect,[34,35] already mentioned previously and shown in Fig. 3(b) and 3(c).

3) Spatially-selective trapping of the second kind of particles (Ag in our case) by the PV field generated during the detrapping process.

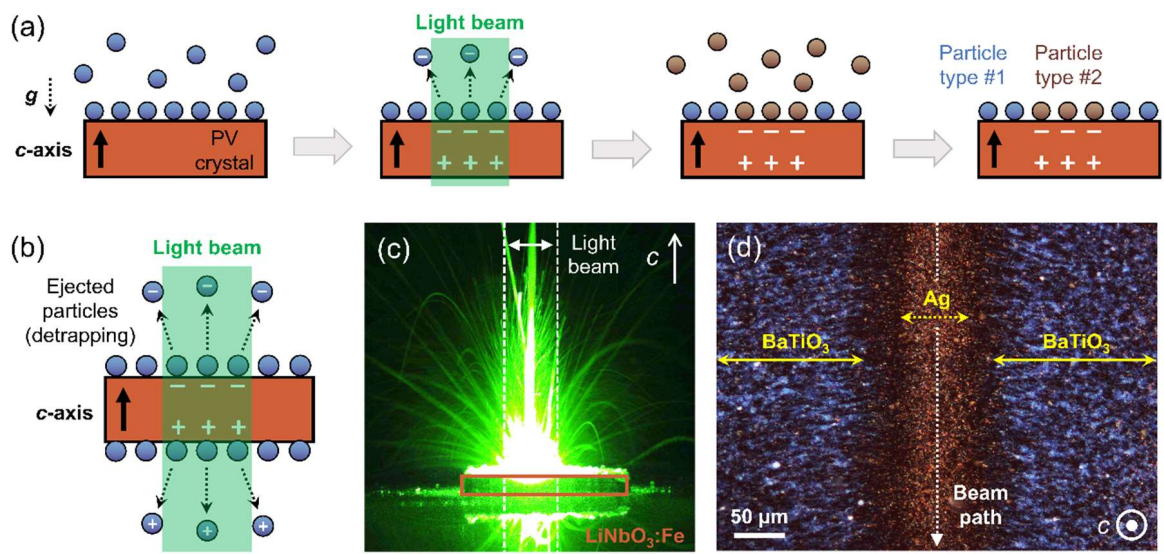

FIG. 3. a) Schematic diagram illustrating the steps involved in the second approach for producing combinatorial patterns. See main text for a detailed explanation of the steps. b) Schematic illustration of the light-driven detrapping process exploited in the second fabrication strategy. c) Example of ejection of Al$_2$O$_3$ particles from a $z$-cut LiNbO$_3$:Fe under illumination. The white dashed line indicates the position of the laser beam (diameter $2w$ = 3 mm, intensity $I$ = 0.3 W/cm$^2$). d) Picture of a combinatorial pattern obtained by using this method: Ag NP stripe surrounded by two regions of trapped dielectric BaTiO$_3$ particles (image taken with crossed polarizers).

A representative combinatorial pattern obtained with this method is shown in Fig. 3(d). It consists of a silver metallic stripe trapped along the laser beam trajectory surrounded by regions of dielectric $BaTiO_3$ particles. A rather clean structure was accomplished in this case, with no evident presence of $BaTiO_3$ residuals at the Ag central stripe and few Ag NPs located at $BaTiO_3$ regions. The origin of such residual contamination is analogous to that of the first approach.

Finally, we propose a third strategy based on *x*-cut substrates that under illumination generate surface charges of both signs, positive and negative, at the active surface (see Fig. 1(a)). Then, if the two types of particles are charged with different signs, they trap on the substrate in the regions with opposite charge sign due to electrophoretic (Coulombian) forces, given by $\mathbf{F_{EP}} = q\mathbf{E_{PV}}$. The shape of these regions is correlated with the light pattern. The successful applicability of this method relies on two aspects:

a) Deterministic control over the charge sign of the particles.

b) High enough amount of charge to ensure that electrophoresis dominates over dielectrophoresis.

Conveniently, the charging mechanism reported in refs. [34,35] provides a simple route to straightforwardly meet both conditions. The steps followed in the third fabrication strategy are illustrated in Fig. 4. It is worth noting that during the whole procedure, all samples are immersed in heptane inside a glass cuvette. First, an *x*-cut crystal (where the combinatorial pattern is going to be produced) is illuminated, hence generating the PV electric field. In our case, we used a static Gaussian beam ($2w = 150$ μm, $I = 6$ W/cm$^2$, exposure time $\Delta t = 1$ min). Then, in the second step, a *z*-cut crystal with a homogeneous background of one type of NPs on the +*c* face is placed next to the *x*-cut crystal (see Fig. 4). When the *z*-cut crystal is illuminated ($2w = 3$ mm, $I = 0.3$ W/cm$^2$, exposure time $\Delta t = 2$ min), the NPs get charged with the same sign as the PV charge, leading to ejection (due to Coulomb repulsion) and electrophoretic trapping on the x-cut crystal. Since particles are negatively charged, they only get trapped on the -c side of the light spot. Finally, this step is repeated with the -c face of another z-cut crystal and a different type of NPs. In that case, the particles get positively charged, and so, they get trapped on the +c side of the light spot.

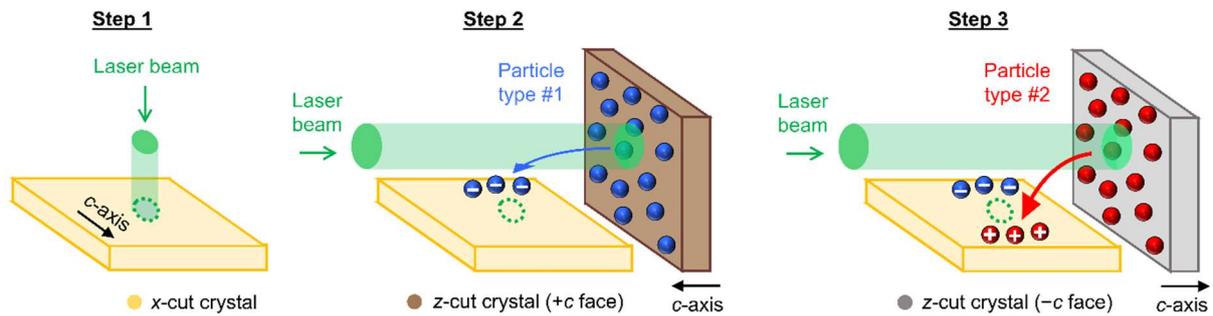

FIG. 4. Sequence of steps for the third fabrication approach, based on the electrophoretic trapping of oppositely charged particles.

First results using this strategy are presented in Fig. 5. Three mixed patterns generated under illumination with a static Gaussian beam are shown. The patterns combine two regions of negative (left) and positive (right) NPs. In Fig. 5(a), the two types of NPs are metallic (Al and Ag), whereas in Fig. 5(b) and 5(c), dielectric (BaTiO$_3$ or Al$_2$O$_3$) and metallic (Ag) particles are combined. It can be appreciated the good quality and homogeneity of the regions with trapped particles, particularly for the metallic ones. Probably because the particles are charged, aggregation effects are minimized for this third strategy compared with the previous ones. Also, the degree of cross-contamination is low and residuals in the middle are almost inexistent, very likely due to the stronger electrophoretic forces. However, this scheme is obviously restricted to binary structures having only two types of particles.

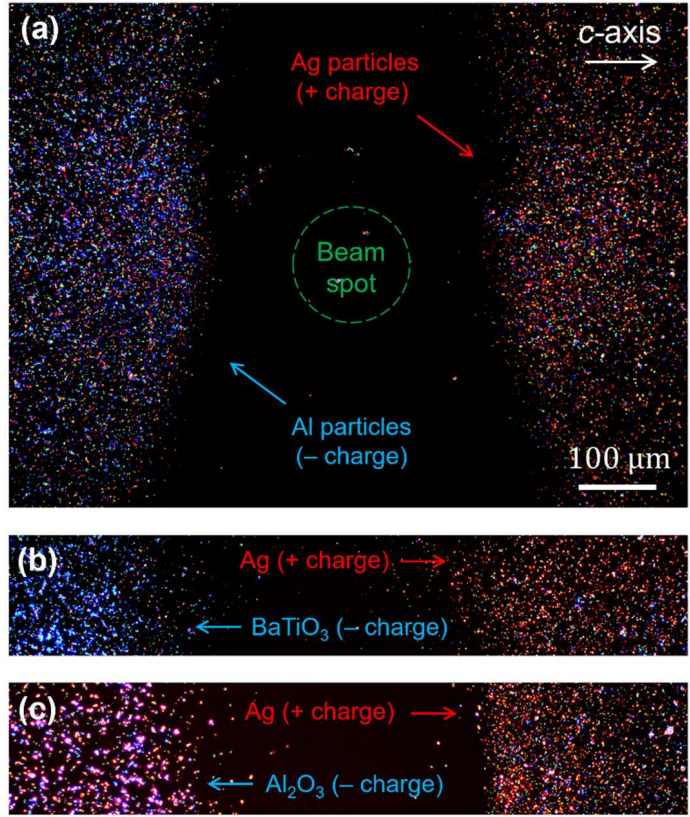

FIG. 5. Binary patterns consisting in two types of oppositely-charged particles trapped on both sides of the illuminated spot along the *c*-axis (captured with crossed polarizers). Positively-charged Ag particles have been used in combination with negatively-charged (a) Al, (b) BaTiO$_3$, and (c) Al$_2$O$_3$ particles.

These results with a simple Gaussian beam demonstrate the working principles of the method, as well as its potential. Further work applying this strategy for more complex and functional structures is in progress. Beyond *x*-cut substrates (typically constrained to 1D patterning), a well-designed synergy between the PV and the pyroelectric effects in *z*-cut crystals can be exploited to generate bipolar surface charges of arbitrary shape.[37] Alternatively, one could also resort to polydomain-engineered substrates.

In conclusion, NP structures combining different types of particles have been successfully assembled using PVOT. Three strategies have been proposed and demonstrated by exploiting the multiple possibilities offered by PVOT for mixed NP patterning. Light-induced trapping and detrapping, positive dielectrophoresis (attractive) and electrophoresis (attractive and repulsive), have been flexibly tailored for that purpose, showing the outstanding versatility of PVOT. Furthermore, all three schemes work satisfactorily with particles of different nature (metallic and dielectric), as well as various sizes. Further efforts in these directions, first introduced in this work, should lead to an optimized cross-contamination and spatial resolution. In the latter case, there is still much room for refinement, ultimately limited by diffraction. Thus, these endeavors should conduct to fabricate improved and more complex combinatorial structures for which a variety of applications is envisaged, such as functional platforms for bio-applications, multiplexed sensing on a single chip or photonic/electronic structures combining metallic and dielectric NPs.

This work has been sponsored by Ministerio de Economía, Industria y Competitividad of Spain under grant MAT2017-83951-R and Ministerio de Ciencia e Innovación of Spain under grant PID2020-116192RB-I00. C. Sebastián-Vicente also gratefully acknowledges financial support through his FPU contract (ref. FPU19/03940).